# Simulating the Phases of the Moon Shortly After Its Formation

**Emil Noordeh and Patrick Hall,** York University, Toronto, ON, Canada
**Matija Cuk,** Carl Sagan Center, SETI Institute, Mountain View, CA

The leading theory for the origin of the Moon is the giant impact hypothesis, in which the Moon was formed out of the debris left over from the collision of a Mars-sized body with the Earth[1]. Soon after its formation, the orbit of the Moon may have been very different than it is today. We have simulated the phases of the Moon in a model for its formation wherein the Moon develops a highly elliptical orbit with its major axis tangential to the Earth's orbit. This note describes these simulations and their pedagogical value.

A recent paper[2] by Cuk & Stewart suggests that the Moon's formation left behind a fast-spinning Earth. During the initial tens of thousands of years after the Moon's formation, the Earth's spin is slowed and its orbit slightly altered by tidal evolution of the Moon's orbit via an "evection resonance". This is a process whereby the Earth's spin decreases through the tidal transfer of angular momentum to the Moon's orbit, which acts to increase the semi-major axis of the Moon's orbit faster than the semi-minor axis, leading to an increase in the Moon's orbital eccentricity as well[3]. Because in an evection resonance the Moon's elliptical orbit around the Earth precesses with a period equal to the Earth's orbital period around the Sun, some of the Moon's orbital angular momentum is transferred to the Earth's orbit around the Sun, causing the semi-major axis of the Earth's orbit around the Sun to increase slightly. This evolution ends when the Moon's motion at perigee becomes nearly synchronous with the Earth's rotation, moving the Moon out of the evection resonance and into the path of increasing orbital semi-major axis it still follows today.

The combination of a Moon much closer to the Earth and in a much more eccentric orbit than it is today would have led to very different phases of the Moon than those with which we are familiar. We follow the evolutionary path of the Moon proposed in Cuk & Stewart that leads to a peak eccentricity of 0.6 and use MATLAB to simulate what the Moon would have looked like along the way[4]. These simulations are publicly available at www.yorku.ca/phall/RESEARCH/MOON/. Snapshots of the simulations with the Moon at perigee and apogee are given in Figure 1. Certain characteristics of the Moon's orbit at various simulated times are given in Table 1. Additionally, the orbits of the Moon around the Earth at these times are plotted in Figure 2.

Initially, 1000 years after formation, the Moon is in a nearly circular orbit with an eccentricity of 0.01. However, the Moon is in very close proximity to the Earth with a semi-major axis that is 12 times smaller than the current semi-major axis. Applied to the moon, Kepler's laws[1] state that:
    1)     The orbit of the Moon is an ellipse with the Earth at one focus.
    2)     A line joining the Moon and the Earth sweeps out equal areas in equal times.
    3)     The square of the Moon's orbital period is proportional to the cube of the semi-major axis of its orbit.
From Kepler's 3rd law we see that a semi-major axis that is 0.08 times the current semi-major axis would result in a Moon with an orbital period that is more than 40 times shorter than it is today. Additionally, with the Moon in closer proximity to the Earth, it would appear in the sky with an angular diameter of 6.25 degrees, as compared to 0.5 degrees today.

By 24,000 years after formation, the eccentricity of the Moon's orbit has peaked at 0.6. This leads to variations in angular diameter from about 2 degrees at apogee to almost 10 degrees at perigee. The Moon would thus oscillate between covering an area 16 times larger to almost 400 times larger on the sky than it does today. In addition, the large eccentricity of the Moon leads to interesting effects such as the time span from full Moon to new Moon being significantly shorter than that from new to full[5].

We have also simulated the Moon at 68,000 years after formation. As seen in Figure 2, this is the point where the Moon makes its closest approach to the Earth, just before leaving the evection resonance.

Finally, we have simulated the current orbit of the Moon alongside the orbit from 24,000 years after formation. In this comparison the dramatic differences in angular size and orbital speed are made extremely clear.

These simulations of a key period in the Moon's evolution have great pedagogical value for introductory physics/astronomy classes. Traditionally abstract exercises in angular size, apparent magnitude (including Earthshine) and orbital dynamics can be linked to the evolution of an astronomical body that is grounded in our experience. For example, a worthwhile challenge would be to simulate the appearance of the Sun and Moon above Earth's horizon over several days at each of the four epochs in Table 1. Additionally, the increased angular size of the Moon in combination with the much shorter orbital period would result in significantly more eclipse events. The simplifications used in the simulations also give rise to several educational questions. For example, the declination range of the Moon is arbitrarily fixed to [-23.5°, 23.5°] for each simulation. Furthermore, in all of our simulations the stars and constellations[6] in the night sky were plotted using data from the Yale Bright Star Catalogue[7]. However, the positions of visual stars in the night sky, and the identities of those stars, would have been entirely different at the time of the Moon's formation billions of years ago.

**Table 1: Characteristics of the Moon's Orbit**

| Time after Formation [years] | Eccentricity | Semi-major axis [$R_\oplus$] | Orbital Period [hours] |
|---|---|---|---|
| 1,000 | 0.01 | 4.9 | 15.2 |
| 24,000 | 0.60 | 8.6 | 35.3 |
| 68,000 | 0.35 | 5.0 | 15.7 |
| 4.5 *billion*[8] | 0.05 | 60.3 | 655.7 |

---

[1] Freedman A. Roger, Robert M. Geller, and William J. Kaufmann. *Universe*. 9th ed. New York: W.H. Freeman and Co., 2011.

[2] Matija Cuk and Sarah Stewart, "Making the Moon from a Fast-Spinning Earth," *Science* 338, 1047-1052 (November 2012).

[3] The angular momentum of an object of mass $m$ orbiting one of mass $M \gg m$ in an orbit with semi-major axis $a$ is $|m\vec{v}\times\vec{r}| = m\sqrt{GMa}$ for a circular orbit, since $|\vec{v}| = \sqrt{GM/a}$ in that case, and $|m\vec{v}\times\vec{r}| = m\sqrt{GMa(1-e^2)}$ for an elliptical orbit.

[4] It is important to note that the times used here are specific to one simulation from Cuk & Stewart. Changing the parameters of the simulation could change the times by a factor of a few.

[5] Cuk & Stewart found an equal probability of perigee being locked at first quarter or third quarter. We have arbitrarily chosen perigee to be locked at third quarter. If perigee was in fact locked at first quarter we would have instead seen the time span from new Moon to full Moon be significantly shorter than that from full to new.

[6] physanim.blogspot.ca/2010/12/fun-with-bright-star-catalog.html

[7] tdc-www.harvard.edu/catalogs/bsc5.html

[8] nssdc.gsfc.nasa.gov/planetary/factsheet/moonfact.html

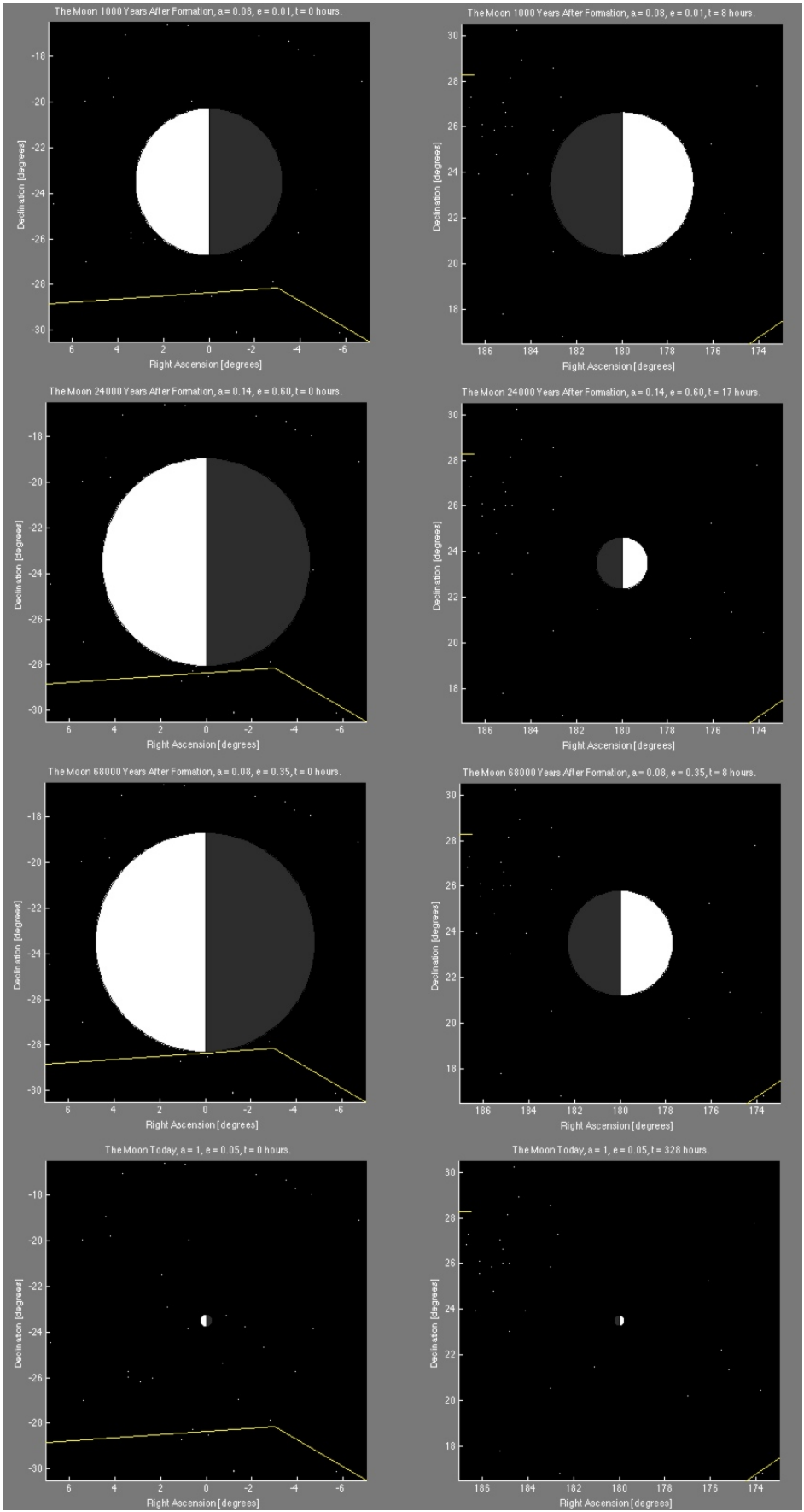

**Figure 1: Snapshots of the Moon at perigee (left) and apogee (right) for each of the simulated points in time. The semi-major axis 'a' is stated in units of the semi-major axis of the Moon's current orbit. The changes in angular diameter over time are obvious.**

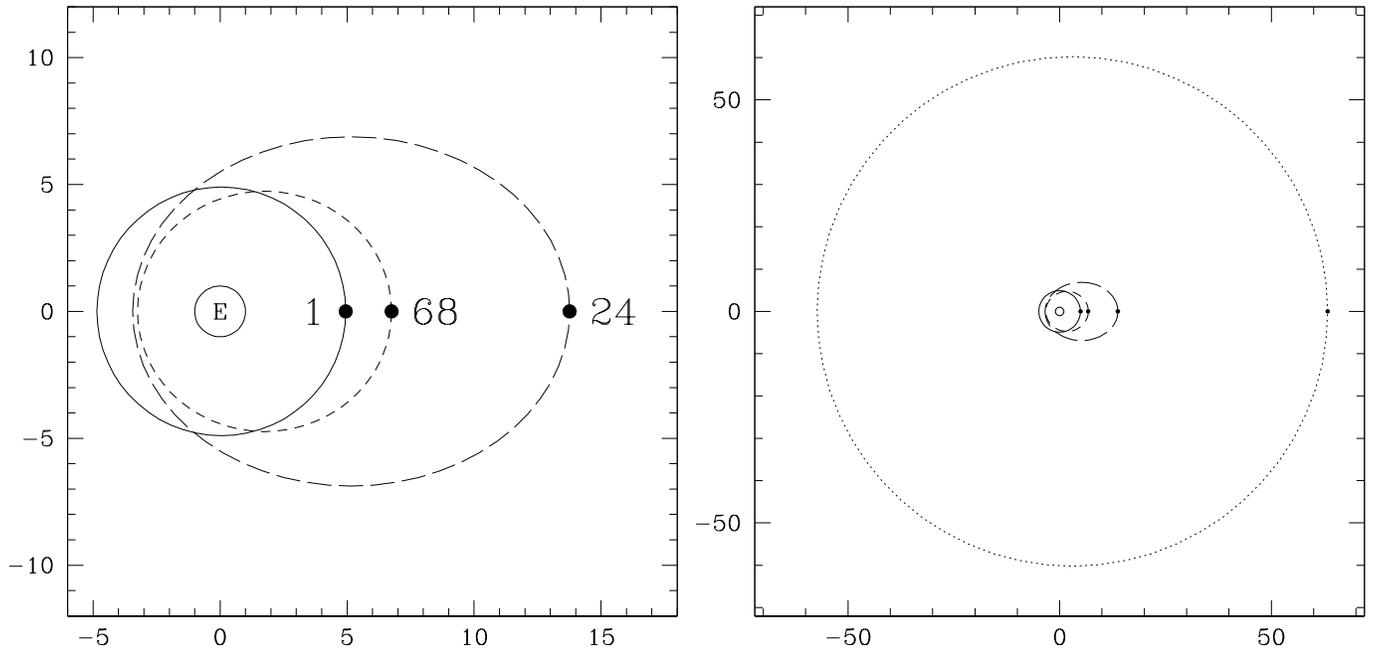

Figure 2: Plots of the orbits of the Moon 1000, 24000, and 68000 years after formation (left) compared to the current orbit of the Moon (right). Axes are in units of Earth radii.